\documentclass[aps,twocolumn,showpacs,superscriptaddress,groupedaddress, nofootinbib]{revtex4-1}

\usepackage{epigraph}
\usepackage{hyperref}
\usepackage{multirow}
\usepackage{mathrsfs,amsmath} 
\usepackage{verbatim}
\usepackage{graphicx}
\usepackage{amsmath}
\usepackage{amsfonts}
\usepackage{amssymb}
\usepackage{amsthm}
\usepackage{bm}
\usepackage{xparse}
\usepackage{xcolor}
\usepackage{textcase}
\usepackage{url}
\usepackage{lipsum}
\usepackage{appendix}
\usepackage{dsfont}
\usepackage{epstopdf}
\usepackage{footnote}
\usepackage{tgbonum}
\usepackage{blindtext}
\usepackage{enumitem}
\usepackage{mathrsfs,amsmath} 
\usepackage{float}
\usepackage{subfloat}
\usepackage{wrapfig}
\usepackage{pslatex}
\usepackage{amsmath}
\usepackage{amsfonts}
\usepackage{setspace}
\usepackage{epstopdf}
\usepackage{wrapfig}
\usepackage{csquotes}
\usepackage{hyperref}
\usepackage{graphicx}
\usepackage{lipsum}
\usepackage{amssymb}
\usepackage{multirow}
\usepackage{xcolor}
\usepackage{framed}
\usepackage{mathtools}
\usepackage{tcolorbox}

\definecolor{shadecolor}{rgb}{0.8,0.9,1}
\DeclareDocumentCommand{\Tr}{m m O{\big}}{{\rm Tr}_{\:\!{#1}}#3({#2}#3)}

\begin{document}
\title{Between understanding and control: Science as a cultural product}
\author{Flavio Del Santo}
\affiliation{Group of Applied Physics, University of Geneva, 1211 Geneva 4, Switzerland; and  Constructor University, Geneva, Switzerland }

\date{\today}

\begin{abstract}
\noindent Since the early days of humankind, people have been asking questions about Nature of two kinds: why did that happen? And how can that be used? In a broad sense, science was born that day. We show indeed that science has two complementary and interdependent souls that aim, respectively, to how to understand and how to control Nature. Through a broad historical analysis, this essay aims to (1) give an account of the development of science as an oscillation and an interplay between its two intrinsic natures, (2) demonstrate that this happened already in ancient times starting from the 6th century BC, and (3) the fact that in different periods one of the two natures was largely favored over the other is a consequence of science being a cultural product of the different social-historical contexts. \end{abstract}

\maketitle

\epigraph{\emph{{Dedicated to the memory of my mentor Angelo Baracca, whose disappearance leaves the world of science and its critics poorer}}}

\begin{figure}[h]
\includegraphics[width=\linewidth]{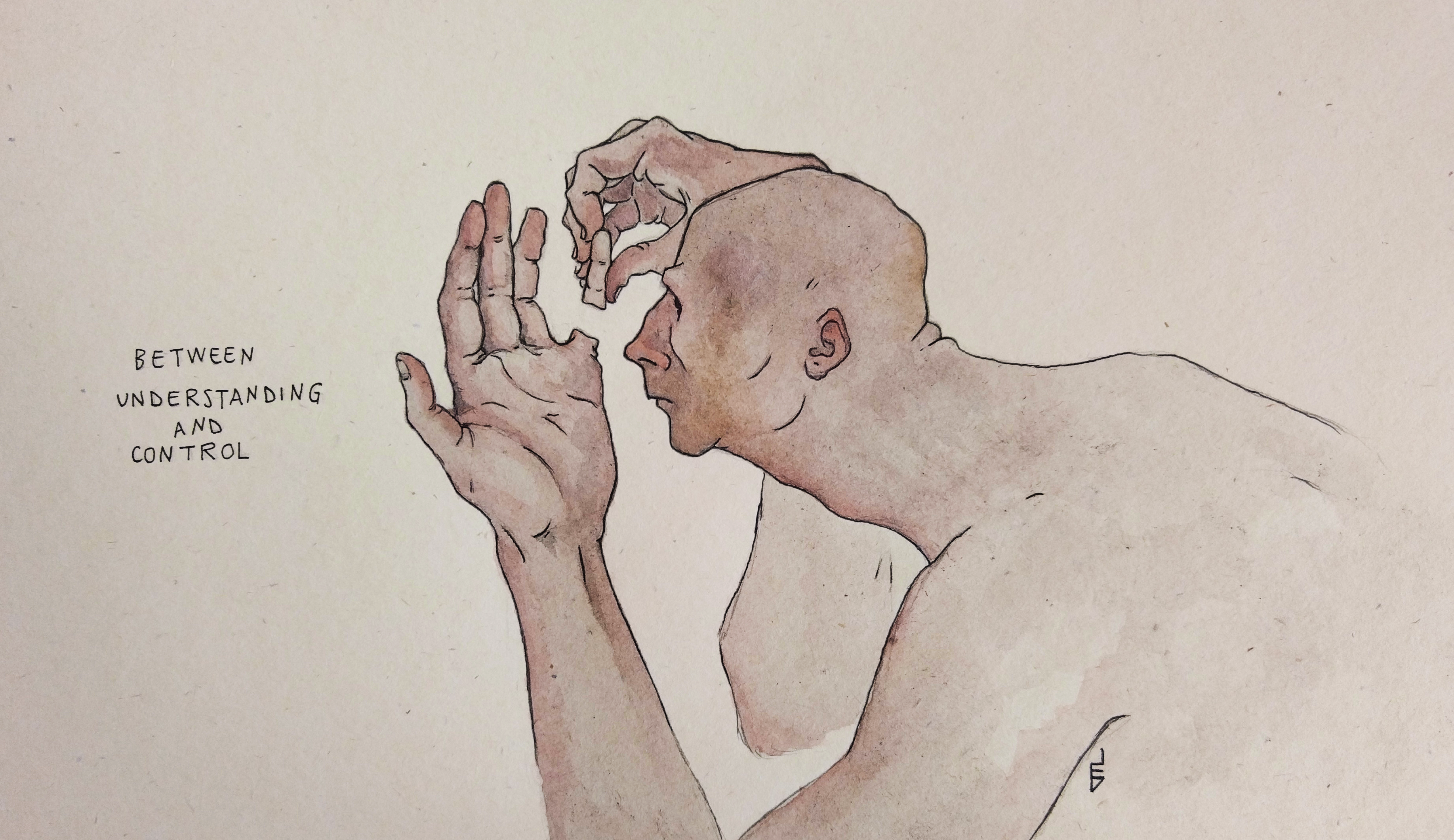}
\caption{Copyright: Jennifer Taufer.}
\end{figure}

\section{The two souls of science}


The word \textit{science} has a very specific meaning in the collective imagination. It invariably brings the thought to the activity that follows the so-called Scientific Revolution of the 16th and 17th centuries, with the pivotal works of Copernicus and Galileo, and the culmination in Newton's theory of mechanics and gravity. But science was born much earlier, both in a broad conceptual sense and, as we shall see, also in what we consider its characteristic collection of systematic methods and specific techniques.\footnote{In what follows we will focus exclusively on natural science, and in particular on Western natural science (with a particular focus on physics), although it is more than likely that these considerations can also apply, at least partly, to other forms of science and to other geo-cultural contexts. We refer the reader, for instance, to the excellent series of books \cite{china} for the development of sciences in China.}

It must have been very early in history---most likely long before some species evolved into \textit{Homo Sapiens}, what we called ourselves today---that somebody started being hungry for explanations. Our world---and we got to learn this even more once we were able to have a glance at what lies beyond what we are able to observe through our bare senses---is extremely complex, full of an overwhelming variety of phenomena: pleasurable or dangerous, wonderful or mysterious. As soon as a minimal amount of cognitive capacities was brought about by the random process of evolution, some species (perhaps some individuals that enjoyed particularly comfortable life conditions that allowed them to indulge in the luxury of thinking in a hostile world) looked around them and started asking certain questions. These questions can be roughly divided into two kinds: why did that happen? And how can that be used? In a broad sense, science was born that day. 

These two types of questions are  associated with the issues of, respectively, how to \textit{understand} and how to \textit{control} Nature.  However, these two different and complementary souls of science have not always been acknowledged, and even if so, their mutual connection was either disregarded or misunderstood. While the explanatory side of science is customarily associated  with its more foundational, speculative, mostly theoretical aspects, the controlling side is considered a direct consequence of the former, in the form of practical-technological applications.

Providing an explanation means to be able to tell a consistent story of how the ``things" we observe came into being, how are they causally related to each other, and how, why and under what circumstances certain phenomena occur. 
This typically requires the introduction of specific languages and practices: on the one hand, one introduces metaphysical (ontological) entities---namely, the theoretical elements, either observable or unobservable, that are required to tell the story (Anaximander's \textit{apeiron},\footnote{With the literal meaning of ``unlimited", ``unbounded", or ``infinite", the apeiron was for Anaximander the principle, a sort of primordial substance,  from which all things generate and eventually go back.}, atoms, fields, forces,  etc.). On the other hand, one may
need to introduce some mathematical language, such as  specific  frameworks that allow to describe and relate the metaphysical entities of a certain scientific theory. The controlling nature of science is more related to the capacity of providing quantitative predictions, isolating systems and phenomena, and therefore being able to replicate the effects of interest on demand.

It is the aim of the present essay to show that (1) science developed as an oscillation and an interplay between its two intrinsic natures, (2) this happened already in ancient times starting from the 6th century BC, and (3) the fact that in different periods one of the two natures was largely favored over the other is a consequence of science being a cultural product of the different social-historical contexts. This reconstruction would allow to make hypothesis on how science could be different.

\section{When was science born?}

\subsection{The pre-Socratics}
Already in prehistoric times the urge for explanation led to the postulation of unobservable elements of reality that would help tell stories about the origin and the relation between natural phenomena. As much as naive religious explanations may sound today (although it seems that they still sound reasonable to an astonishingly large part of the human population), at an early stage of knowledge they represented a powerful tool for the elevation of human thought. Ascribing the alternation of night and day, the thunderbolt, the tumultuous ocean, the transformation of water in vapor, etc. to specific deities was a major step towards the explanatory nature of science.

Early civilizations, remarkably in the Fertile Crescent,
started to carry out systematic observations, giving quite impressive contributions to science, especially in  astronomy and in medicine \cite{lindberg}. But it was likely in Ancient Greece that the explanations of \textit{how Nature does it} started acquiring what we consider today a scientific character. As early as in the 6th and 5th century BC, the Ionian School (Thales, Anaximander, Anaximenes), Heraclitus, Anaxagoras, the Eleatic School (Parmenides, Zeno, Empedocles), and Democritus approached the explanation of natural phenomena by constructing causal theoretical hypotheses---involving observable (e.g., water, or fire) or unobservable entities (e.g., the \textit{apeiron}, atoms, or seeds)---aimed at unifying natural phenomena by means of 
\begin{itemize}
    \item [(i)] general principles (if it was only one, the \textit{arche}, one speaks about ``material monism"), that did not involve sentient supernatural beings (deities), and
    \item  [(ii)] introducing economical ontologies (in terms of the amount of postulated metaphysical entities required for an explanation to be consistent).
\end{itemize}

A prime example of this new scientific  approach of the explanatory kind was the one  put forth by Anaximander of Miletus. In what is apparently the only survived excerpt of his writings (relayed by Simplicius) one reads: ``The things that are perish into the things from which they come to be, according to necessity, for they pay penalty and retribution to each other for their injustice in accordance with the ordering of time" \cite{curd}. In his book ``Anaximander and the birth of science" \cite{rovelli}, C. Rovelli already rightly noticed that this contains the first testimony of the idea of a natural law, a prescription that governs things by necessity while they evolve in time. This is an exceptionally powerful abstraction, and although it does not yet contain any form of prediction, it sets the structure for what is still today believed to be a scientific explanation.

In an excellent while not very well-known essay, ``Back to the pre-Socratics", K. R. Popper contends that modern science has forgotten its pivotal role of explanation by moving the focus on collecting and analysing data, under the push of an empiricist and positivistic tradition. However, he maintains, ``Western science---and there seems to be no other---did not start with collecting observations [...], but with bold theories about the world." \cite{popper}. Popper argues that Anaximander's idea that earth is suspended in space and that its stability is ensured by symmetry arguments (i.e., that there is no privileged direction onto which to fall) is not provided by observation but by reasoning, making it ``one of the boldest, most revolutionary, and most portentous ideas in the whole history of human thought."   \cite{popper}. This is a striking example of ``good science", not because Anaximander's theory turned out to be correct, but because it provided a consistent explanation, scientific to the extent that completely fulfills the two points (i) and (ii) above. What Popper involuntarily acknowledges here is the distinction between the two souls of science we have previously described.
In contending that we should go back to the pre-Socratics, Popper unwittingly supports the idea that science should go back to its more explanatory nature rather than its  controlling nature. What Popper fails to realise, however, is, on the one hand,  that these two natures complement each other, and, on the other, that this separation of the two natures of science did not happen accidentally but was rather a process influenced and even guided by the development of (Western) society. Science is a human activity and, as such, a cultural product that developed in its particular socially interested and oriented way. The propensity towards a larger appreciation of either of the two natures of science should hence be studied in its cultural context.

\subsection{The Hellenistic Age}

Before elaborating in detail on the social-cultural development of science as the engine for its oscillation between its two intrinsic natures, we shall still remain within ancient history for a little longer. In fact, one may still think that the considerations made about the pre-Socratics only show some partial anticipation of what science is supposed to be. Instead, in what has become a quite influential essay, ``The forgotten revolution"  \cite{russo}, L. Russo maintains, with detailed and sound arguments, that science---basically in its current, modern conception and methodology---was an already well-established practice in Hellenistic times (sometimes identified as the period between 323 BC, the year of death of Alexander the Great, and 415 AD, when the mathematician Hypatia of Alexandria was lynched by a mob of Christians). Hellenistic thinkers, who had Alexandria of Egypt as their main center, were not just natural philosophers who anticipated some aspects of modern science, but rather  full-fledged scientist according to modern methodological standards: 
\begin{quote}
    If an essential characteristic of the experimental method lies in making quantitative measurements, the systematic use of such measurements had been present for many centuries in astronomy [...]. In the early Hellenistic period quantitative measurements were extended not only to fields such as mechanics and optics, but to the medical and biological sciences [...].
    
If by experimental method we understand the practice of observation
under artificially created conditions, the most significant examples are perhaps in pneumatics, where we see the systematic construction of experimental gadgets for demonstrations, but examples are documented in other areas as well. \cite{russo}
\end{quote}

In the Hellenistic period, in Alexandria, worked and lived Euclid, author of the most read manual of geometry of all times, Ctesibius, who invented pneumatics and established a school of mechanics, and Aristarchus of Samos, who was the first to propose a heliocentric model. Moreover, Eratosthenes, head of the Library at Alexandria, was the first to measure the circumference of the Earth with an impressively clever experiment (with an error on the real value of less that 3\% \cite{russo}), whereas Apollonius of Perga developed the theory of conic sections, and Hipparchus of Nicaea, who is considered the greatest astronomer of antiquity (he was the first to introduce the epicycle model).
A particular attention is deserved by the work of Archimedes of Syracuse---who most likely studied in Alexandria in the third century BC, and kept a regular correspondence with Alexandrian scientists---whose studies on statics, applied geometry and hydrostatics would have been unparalleled for centuries to come. His work was based on systematic methods of practical experimentation and advanced mathematical techniques. Worth recalling are his findings on the   buoyant force exerted on a body immersed in a fluid, which he (experimentally) demonstrated to be equal to the weight of the fluid displaced by the body (\textit{Archimedes's principle}), his approximation of the value of $\pi$, and the systematic use of the method of exhaustion---previously introduced by Eudoxus of Cnidus\footnote{Eudoxus, one of the most important mathematicians of the Classic Age, is remembered for having been the first to propose a (quantitative) geometrical model of the motion of celestial objects, then adopted by Aristotle.}---for the calculation of areas and volumes which anticipated definite integrals in mathematical analysis \cite{rufini}.\footnote{The rigorous use of the method of exhaustion as well as the method of mechanical theorems are expounded in Archimedes' treatise ``The method", fortuitously rediscovered in a palimpsest in 1906 by J. L. Heiberg, then lost again and retrieved only in 1998.} 
 Finally, in the 2nd century AD, Ptolemy was active in Alexandria where he wrote the ``Almagest", a treatise that raffinò the geocentric cosmology which became  the canonical model explanation of the observed universe until the acceptance of Copernicus' heliocentrism. 
 
We ought to notice that this ``forgotten scientific revolution" that boomed during the Hellenistic period strikes us as very modern mostly because it makes use of empirical observations, reproducibility (not systematic, but to the extent that experiment were carried out in artificially created conditions), and mathematical modelling. All of these being characteristics that we attribute to the ``proper science" that arose after the Scientific Revolution of the 16th-17th centuries. While this is surely true, it should be noticed that this kind of science is clearly more, althoutgh not exclusively, of the controlling type, rather than of the one aimed at understating and explaining Nature.

We have therefore shown that already in ancient times Western science had developed in a certain period (pre-Socratic) as a doctrine to understand nature, and in a later period (Hellenism) more, although not exclusively, as a quantitative empirical activity, that could be used also to control nature and produce new technology. 

We want to stress here that we are not proposing a hierarchy between these two intrinsic natures of science, following the quite naive narrative that there is a pure, ``sacred" science of explanation that then gets vulgarized by down-to-earth applications (the controlling side of science). Rather, we identify these two souls of science as two complementary characteristics. 
We will discuss in the next section how science developed through the interplay between understanding and control.

The Hellenistic scientific ``golden age" slowly declined with the rise of the Roman Empire, without  necessarily being caused by the latter, although one of the most dramatic events in the history of human culture happened at the hand of Julius Caesar, whose army set on fire the  Library of Alexandria during civil war of 48 BC \cite{plutarch}.\footnote{It seems that the library was only partly burned in that occasion and, while subsequently rebuilt, it saw a gradual decline  in the following Roman period, as well as other fires that ultimately destroyed it.} This probably dissolved a considerable part of the academic and scientific knowledge of humankind at that time, a loss for culture that can only be compared today to the destruction of the Internet without any local backups. While known for its military power, its contributions to architecture, law and politics, Ancient ``Rome [has been] a civilization to which science remained foreign." \cite{russo}; this was also stressed by C. Boyer in his ``A history of mathematics", wherein he bluntly maintains that Cicero's discovery of Archimedes' tomb has been ``almost the only contribution of a Roman to the history of mathematics"  \cite{boyer}. And we will not dwell on the subsequent centuries of obscurantism and bigotry of the Middle ages, that---despite recent trends that would like to rehabilitate those times---definitely did not see any notable contributions to the development of science if not, perhaps, by stimulating its renaissance by a repudiation of the Medieval values.

What we conclude from this glimpse at ancient history is that  science has indeed been different, but this has happened in a much less linear way than the popular tradition has portrayed it. The standard story, in fact, is that there has been several centuries of pre-scientific knowledge in the form of natural philosophy, astronomical observations, technological achievements in mechanics and navigation, that set the stage for what matured into proper science after the Renaissance. The latter period is supposed to have marked a clear cut by defining the standards and methods of science and have introduced us into an actual scientific era that has remained more or less constant ever after.

The view that we propose here, on the contrary, acknowledges that science was present in human civilization all along and that the way it has been different should be sought in which of its two roles science has from time to time acquired. Namely, by the oscillation between its more explanatory role and its more controlling role on Nature. We have recalled that in the pre-Socratic times science has developed almost exclusively in the direction of understanding. On the other hand, in the Hellenistic period, which was exceptionally prolific for science, this activity drifted more towards models and experiments, producing at the same time the mathematical and the technical tools that would allow the development of (quantitative) practical applications, i.e., a control on the natural phenomena.
The Scientific Revolution of the 16th and 17th centuries have perhaps the merit of having considered more deeply the interplay between the two natures of science and while the progress of the study of movement of falling bodies, astronomy, optics---carried out by Bacon, Locke, Galileo, Descartes, Huygens, Boyle, Hooke, Torricelli, etc.---leaned more towards applications and therefore control, Copernicus' new cosmological system and Newton's theory of mechanics and gravitation have 
 a seemingly more general scope of providing explanations.

If one is to accept the distinction here proposed, it begs the question, what has \textit{caused} science to oscillate between its two different souls? In the next section, we will analyse in greater detail the development of what has been called ``modern science" (i.e. starting from the Scientific Revolution and its broader establishment in the subsequent Age of Enlightenment) in its cultural context.

\section{On the cultural context of science:\\Could have science been different?}

Since the early 1930s, historiography of science started abandoning the naive positions of the Enlightenment according to which scientific knowledge is a linear accumulation of (approximated) truths, with its romantic and idealized narrative of rationality, objectivity, freedom from biases and preconceptions. On the other hand, science started being recognised as a human activity. In fact, a new trend appeared---which became known as \textit{externalism}---with the focus on how external factors in the social and cultural context influence, guide, and even determine the evolution of science.

After the pioneering work of Soviet physicist and historian B. Hessen on the social-economical factors that influenced the work of Newton \cite{hessen} (see further), L. Fleck explained the progress of science as an agreement between cultural circles that he named ``thought collectives" (\textit{Denkkollektive}) \cite{fleck}, which in turn influenced T. Kuhn's celebrated book ``The structure of scientific revolutions" \cite{kuhn}. Meanwhile, R. K. Merton developed further the approach of Hessen \cite{merton}, becoming in the 1970s the father of what is known as ``sociology of science" \cite{merton2}; this eventually evolved into the modern academic discipline of ``science and technology studies" \cite{sts} that deals with the historical development and the consequences of science and technology in their cultural and social contexts. Such an externalist approach to the historiography of science is what will help us analyse some notable historical periods and try to find patterns  that justify the oscillation of science between its explanatory and more controlling sides.

\subsection{The bourgeois Scientific Revolution of Newton} 

In the midst of the societal changes that led to the Scientific Revolution, most of the scientists the likes of Galileo focused on experimentation on isolated systems in artificially induced initial conditions, such that the sought effect could be triggered on demand, leading to a science aimed to have control over natural phenomena. In general, the empiricist tradition that can be traced back to F. Bacon aimed ``to create
a method for \textit{controlling} the forces of nature." \cite{hessen} (The emphasis is ours). However, despite his generally empiricist approach, Galileo gave essential contributions  to the pure understanding of Nature. Although his name is often associated with heliocentrism, to the extent that his phrase  ``And yet it moves!" entered  popular culture, the heliocentric theory was (re)introduced by Copernicus and subsequently improved by Kepler (who introduced the elliptical orbits) and Galileo, who provided some observational but not crucial support to that theory (as a matter of fact, he falsified the geocentric cosmological system of Ptolemy with his observations of the phases of Venus). However, one of the most important conceptual discoveries---surely not stressed enough when speaking about Galileo's achievements---was the understanding, based on his observations with the telescope, of the Earth-like nature of the Moon; namely, the presence of ``imperfections" in the form of mountains and craters. This falsified the explanation given by Aristotle, which had been accepted and never questioned for almost two millennia, that the celestial objects were perfect ethereal spheres, thereby putting into a profound crisis the whole tenability of the Aristotelian physics.     

What is regarded to be the scientific explanation \textit{par excellence}, however, is the theory of mechanics and gravitation  put forward by Newton in 1687 in his ``\textit{Principia}" \cite{newton}. It so seems that Newton's science is of a purely theoretical type, aimed exclusively at the explanation of the motion of bodies under the influence of mechanical and gravitational forces as crystallized by ``the traditional representation of Newton in the literature as an Olympian standing high above all the `terrestrial' technical and economic interests of his time, and soaring only in the lofty realm of abstract thought." \cite{hessen}.

However, Hessen carried out a Marxist analysis of the social and cultural context in which Newton's work was developed. Therein, he convincingly shows that ``despite the abstract mathematical character of exposition adopted in the \textit{Principia}, not only was Newton by no means a learned scholastic divorced from life, but he firmly stood at the centre of the physical and technical problems and interests of his time." \cite{hessen}. Indeed, the times when Newton carried out his work correspond to the English Civil War, the establishment of the Commonwealth and the struggle between the rising bourgeoisie---to which Newton belonged---against the feudalism (that had still the support of the intellectuals in the  universities born in the Middle Ages). At that time, manufacture was making its entrance on the economic landscape and the merchant capital was becoming the predominant economic force. The technical demands of this rising economical system and reorganisation of society, of which we have evidence that Newton was well-aware, dealt mainly with land and marine transport, heavy (mining and metallurgical) industry, and military technologies. All of these problems are related to technical issues of mechanics, that indeed became the focus of the study of physics in the 17th century, namely the problem of simple machines, the free fall of bodies and the trajectory of projectiles (ballistics), problems of celestial mechanics (also related to systems of orientation in open seas), problems of 
 hydrostatics and aerostatics (related to navigation and again ballisitcs). The only problems of war industry, civil industry, and commerce that were not of mechanical nature were the ones related to the actual production of firearms and industrial machines, which are
problems of metallurgy, of which however Newton was a world-class expert due to his deep interest in alchemy that was grant him a post the Royal Mint. 
About the period in which Newton was active, F. Engels concludes that ``step by step, science flourished along with the bourgeoisie. In order to develop its industry, the bourgeoisie required a science that would investigate the properties of material bodies and the manifestations of the forces of nature." (quoted in Ref.\cite{hessen}). Therefore, it is possible to argue that the explanatory science of Newton was a consequence of the demands of the society of his times, which required a more thorough understanding of the phenomena that had been able to only partially control.  

\subsection{Industrial revolutions and a new science}

The trust in science that followed the success of Newton's theory---that indeed allowed to solve \textit{any} problem of mechanics and hence to design any possible technology based on that field---was overwhelming and led to the Age of Enlightenment. Society became not only more aware and supportive of science and its values, but also dependent on science for the development of its infrastructures and means of production. Industry thrived in the UK (followed closely by continental Europe and the US) with the invention of the steam engine--- by scientists and engineers like T. Savery and J. Watt---and mechanization completely  superseded  manufacturing: The Industrial Revolution broke out at the turn of the 19th century. Science had never been so important for society, because never was the control exerted by humans on Nature so strong. 

A. Baracca, S. Ruffo and A. Russo, in the essay  ``Scienza e industria 1848-1915"---which has regrettably never been translated into English and has recently been called ``an unjustly forgotten treasure of the history of science" (see J. Renn's preface to \cite{baracca})---have reconstructed the complex interplay between science and industry during the Second Industrial Revolution in the social-economical context of the rise of the modern capitalistic society. They notice that until the  mid-19th century,

\begin{quote}
the new ``bourgeois" science, setting itself the task of quantitatively \textit{controlling} the processes that underlay the new technologies, explicitly declared the need to limit itself to the facts of experience and programmatically rejected  to use hypotheses of a metaphysical nature. [...] One would regard empirical facts with the explicit purpose of expressing laws in mathematical form in order to \textit{understand} and \textit{control} them, without using any hypotheses that were not immediately and directly testable. \cite{baracca} [Emphases are ours].
\end{quote}

The thesis of the book is that in the second half of the 19th century the phenomenological approach of positivistic science gives away to a new method strongly rooted in the use of mathematical models and  theoretical-metaphysical hypotheses. This was not because science was experiencing an internal crisis, but rather because the dynamism of the new system of industrial production and enterprise required to overcome the mere empirical approach in order to  suggest unexpected connections between phenomena and different fields that could lead to original practical applications. 

Indeed, the demands of the capitalistic society become more pressing. Already by the mid-19th century, the coke had substituted the charcoal for the fusion of metals in Great Britain, followed by Belgium, France and Germany. The techniques of production of steel underwent a rapid improvement and the \textit{Bessemer process} (invented in 1856) considerably enhanced its efficiency: around the 1870s the annual world production of steel amounted to around 500000 tons (half of which in the UK alone). The textile industry underwent a vast	mechanization, and the American I. M. Singer invented the sewing machine in 1851. The chemical industry thrived: potassium carbonate and sodium carbonate (especially thanks to the new \textit{Solvay process})---both involved in the bleaching of textiles and in the production of soap, glass, and gunpowder---started to be produced in massive quantities. Germany, which had remained quite behind in the First Industrial Revolution, rapidly stood  out as \textit{the} new industrial powerhouse, equating the British production of steel around 1890 and doubling it after 1910; Germany also introduced specialised education in the form of Polytechnics, and its chemical industry grew immensely in the following decades (the companies Bayer and BASF were founded in 1863 and 1865, respectively) with the involvement of scientists as leading positions.

It is in this period that science experienced a new turning point. In 1865, J. C. Maxwell published the work ``A dynamical theory of the electromagnetic field" \cite{maxwell}, wherein he unified the electric and magnetic fields (which had  already separately  been the subject of extensive studies, both theoretically and for applications) into a single ``electromagnetic wave". He also found out that these waves travel at the speed of light, leading to the understanding that light itself is an electromagnetic wave. But this also implies that there must be other waves of different frequencies, such as the at that time unobserved radio waves, which were indeed firstly predicted by Maxwell's theory (and experimentally confirmed by H Hertz in 1886). 

Moreover, the works of R. Clausius, L. Boltzmann and again Maxwell, led to formulate the so-called ``Kinetic theory of gases" which postulates that gases are composed of identical microscopic particles (atoms, molecules) moving in rapid motion and that undergo random collisions between each other. This allowed to explain the macroscopic properties of gases, such as temperature, pressure, and volume. Moreover, soon after, this theory evolved into statistical mechanics, mostly thanks to Boltzmann  and J. W. Gibbs, which allowed to explain thermodynamics---developed as a heuristic theory for heat machines in the previous century---again in terms of statistical considerations on ensembles of microscopic particles, thereby reducing thermodynamics to mechanics at the conceptual level.

We should stress here that what electromagnetism and statistical mechanics have in common is that they are theories developed  to explain known but not fully understood phenomena (again, in the sense that there was no way to tell a consistent story that would causally relate the known effects). This required once more to introduce metaphysical unobserved entities: The (mostly) invisible electromagnetic waves and the microscopic atoms. However,
\begin{quote}
    this is by no means a return to metaphysical, unverifiable and arbitrary conjectures. Hypotheses and models, of whose arbitrariness one is fully aware, are now taken as natural conjectures that transcend mere empirical facts;  but they are, exactly for this reason,  capable of "predicting" new orders of phenomena or unsuspected connections between them, susceptible, however, to experimental verification and thus capable of pointing to new paths for the development of the productive forces. \cite{baracca}.\footnote{This can be identified as a \textit{progressive research programme} as expounded by I. Lakatos \cite{lakatos}.}
\end{quote}

Note that, to take a step outside of physics, around the same years, C. Darwin was formulating his theory of evolution based on natural selection, published in his ``On the origin of species" in 1859 \cite{darwin}. This is another example of science aimed at understanding rather than controlling. In fact, this work marked a turning point with respect to the previous studies (the first fully developed theory of evolution was put forward by J.-B. Lamark in 1809). Its main novelties, and what caused it a strong opposition, were, on the one hand, the abandonment of a teleological explanation replaced by a mechanism of selection that draws paths within random fluctuations (i.e., mutations). On the other hand, and more importantly for our analysis, Darwin adopted an approach based on theoretical conjectures rather than on directly observed facts, similarly to Maxwell's and Boltzmann's methods in physics. In fact, while it is undoubtful that Darwin was a field researcher who had evidence of different lines of evolution from observing existing species, the explanation in terms of a mechanism (survival of the fittest) that selects certain random mutations in relation to their environment was an original theoretical hypothesis. And this aspect is remarked by Darwin himself in a letter to a colleague immediately after the publication of his book: ``What you hint at generally is very, very true: that my work will be grievously hypothetical, and large parts by no means worthy of being called induction from too few facts." \cite{letters}.

As much as the new approaches shared by Maxwell, Boltzmann, Gibbs, and Darwin  marked a clear cut from the positivistic past---insofar as they adopted bold theoretical hypotheses and introduced metaphysical unobserved entities aimed at explaining the observed phenomena---they suffered the limit of staying tied to an old mechanistic paradigm of explanation. Both statistical physics and Darwin's evolutionism introduced randomness into the natural sciences, but as mere working hypothesis and did not dare to bring it to a more fundamental level. Baracca \textit{et al.} \cite{baracca} argue that at the turn of the 20th century, especially in Germany, the opening of new spaces at the productive and social levels pushed the search for even bolder and more creative scientific practices, that would go beyond the mechanistic view.

To summarize, science in the Age of Enlightenment up until the First Industrial Revolution was carried out in the most controlling way over Nature, avoiding any metaphysical assumption and sticking merely to the empirical facts and their applications. Interestingly enough, the Second Industrial Revolution saw such a rapid and uncontrolled growth of the demands of the industrial society that science had to become more daring. This led to a renaissance of a more explanatory science that introduced bold theoretical and metaphysical hypotheses to explore as more as possible the spectrum of natural phenomena and try to induce unforeseeable applications. We here see the complementarity of these two sides of science, and the complexity of their interplay, to their full potential.


\subsection{Could have quantum mechanics been different?}

The spark of the new physics that revolutionised the landscape at the turn of the 20th century was ignited by M. Planck, quite symbolically in the year 1900. He adopted the same kind of theoretical expedient used by Maxwell and Boltzmann that we have already discussed in the previous section, but somehow Planck dared more. To explain the observed energy spectrum of a what is called a black body (i.e., an idealised physical system that absorbs all the incident electromagnetic radiation), Planck assumed that energy can only be exchanged in discrete packages, namely that there is a minimal amount, an  ``atom" of electromagnetic radiation, despite the fact that Maxwell's theory describes light as a continuous wave. Although he was to win the Nobel Prize for his bold hypothesis, Planck never fully accepted it as a new fundamental discovery about Nature, but remained essentially tied to the idea that his assumption was somehow a clever mathematical strategem that could be eventually explained away in mechanistic terms, a tradition coming from the the previous century.

The following years saw an incredibly prolific series of successful attempts to apply Planck's discretization hypothesis, named the \textit{quantum}\footnote{A quantum is the minimal amount of a physical quantity.} hypothesis, which eventually evolved into quantum theory. Notably, this was  used by A. Einstein to explain the photoelettric effect (the phenomenon for which metals eject electrons when hit by electromagnetic radiation), in the revolutionary paper that for the first time interpreted the quantum of light (\textit{photon}) as a real physical entity (contra Planck). Moreover,  the quantum hypothesis allowed N. Bohr to explain the internal structure of the atom.\footnote{We will not discuss here in any more detail the ferment in physics at the turn of the 20th century, that has been the object of thorough study. We refer the interested reader to, e.g., M. Jammer's work \cite{jammer} and references therein. Moreover, we will shamelessly omit here the development of the other revolutionary theory that developed around the same time, namely relativity.} Physics thus underwent a period of exciting experimentation with new theoretical attempts, but it soon became manifest that the mere introduction of some metaphysical hypotheses, but still fully  within the domain of mechanistic explanations, was no more sufficient. A deep crisis outburst.    

As Plank himself stated in a famous quote, ``a new scientific truth does not triumph by convincing its opponents and making them see the light, but rather because its opponents eventually die and a new generation grows up that is familiar with it." \cite{planck}; a rather ironic remark, since after igniting the revolution he remained a leader of  the ``conservative faction" for the rest of his life. In fact, it was a new generation of very young physicists---encompassing W. Heisenberg, W. Pauli, P. Dirac, together with the less young M. Born and E. Schr\"odinger---that lay the groundwork for the new quantum theory in the mid-1920s. Quantum physics, while being extremely successful for predictions (more than any other theory ever formulated) shattered many of the mechanistic beliefs that had characterized the form of explanation of the previous physics (which, by contrast, is now called classical physics). It abandons the idea of causal determinism, i.e. that the laws of physics determine uniquely the past and the future evolution of physical objects (and ultimately of our universe), providing only fundamentally probabilistic predictions. It introduces discreteness over classical continuity, the impossibility of measuring certain pairs of physical variables at the same time (\textit{Heisenberg uncertainty principle}), and forces us to rethink our idea of locality (i.e., that systems cannot influence each other at-a-distance). And it even questions the existence of an objective reality independent of an observer. Note that these positions, that are still the object of debate to date, are really radical and open science to a kind of explanation that would have not been considered scientific in the previous centuries. 

Continuing on the route that we have been following in previous sections, it is interesting to ask whether it is possible to identify cultural causes that led to the particular kind of explanation offered by quantum physics. This is by no means intended to scale down the fact that there were objective physical problems that begged for explanation at the end of the 19th century. The question here is whether the specific type of radical explanation offered is related to its social and cultural context. In what has by now become a classic on the history and philosophy of quantum theory, ``The conceptual development of quantum mechanics", M. Jammer hits at the fact that ``certain philosophical ideas of the late nineteenth century not only prepared the intellectual climate for, but contributed decisively to, the formation of the new conceptions of the modern quantum theory [...] contingentism, existentialism, pragmatism, and logical empiricism, rose in reaction to traditional rationalism and conventional metaphysics." \cite{jammer}.

These ideas were thoroughly developed by P. Forman in the early 1970s, in an influential work  that has become known in history of science as the (first) ``Forman thesis"  \cite{forman}. The latter reconstructs the development on quantum theory in the cultural context of the Weimar Republic (1918-1933) in Germany, which saw a particularly ``hostile intellectual environment" for standard scientific explanation, with the growth of a particularly  strong opposition of rationalist and causal-realist approaches:
\begin{quote}
   [I]n the aftermath of Germany's defeat [in World War I] the  dominant intellectual tendency in the Weimar academic world was  a neo-romantic, existentialist ``philosophy of life," reveling in crises and characterized by antagonism toward analytical rationality generally and toward the exact sciences and their technical applications  particularly. [...]
 
  There was in  fact a strong tendency among German physicists and mathematicians  to reshape their own ideology toward congruence with the values and  mood of that environment---a repudiation of positivist conceptions  of the nature of science, of utilitarian justifications of the pursuit of  science, and, in some cases, of the very possibility and value of the  scientific enterprise. [...]

  [T]he movement to dispense  with causality in physics, which sprang up so suddenly and blossomed so luxuriantly in Germany after 1918, was primarily an effort  by German physicists to adapt the content of their science to the  values of their intellectual environment. \cite{forman}.
\end{quote}

The Forman thesis became very influential in the history of modern physics, paving the way to investigations that reconstructed the social and cultural context leading to the so-called period of “shut up and calculate" (see e.g., \cite{kaiser}) after World War II and the “successful" enterprise that led to the atomic bomb (if one can talk of success when referring to such a use of science for a device that can annihilate humankind). Here the economic and social factors are very manifest in the deliberate political decisions of Western countries (with the US at the forefront) that aimed to channel the work of scientists, and physicists in particular, towards practical applications, often of a military nature. As a testimony of the shut up and calculate culture, in 1951, the words of a report written by a leading member of the US Atomic Energy Commission are remarkable, when he referred to physicists as a
“war commodity”, a “tool of war”, and a “major war asset” to
be “stockpiled” and “rationed” (quoted from \cite{kaiser}). This period is one of the most explicitly oriented towards control rather than understanding of natural phenomena, with an active policy to guarantee that no resources would be “wasted" on foundational (i.e. explanatory) science. 

Fortunately, starting with the pivotal and for long unappreciated result of J. Bell in 1964---an inequality that allows to experimentally rule out the possibility of explaining quantum theory by adding local hidden variables (\textit{Bell Inequality} \cite{bell})---quantum foundations, whose aim is primarily to find explanations of phenomena in the quantum domain, were slowly revived. In the 1970s and 1980s, a few pockets of resistance countered the “shut up and calculate" trend and led to the field of modern foundations of quantum mechanics, which in turn paved the way for quantum information science. This happened in other interesting cultural contexts, such as a radical leftist critique of science in Italy and France (see \cite{freire, io}), the hippie counterculture in the US \cite{kaiser}, and a movement of reconciliation of physics with philosophy in Postwar Vienna \cite{ioemanu} (for other “Places and Contexts" relevant to the rebirth of quantum foundations, see  Ref. \cite{oxford}, Part III). Quantum foundations eventually made it (again) into mainstream physics, as witnessed by this year's Nobel Prize in Physics which was awarded precisely for the violation of Bell inequalities. 

In conclusion, we have used the historical development of quantum mechanics as a case study to show that the oscillation between a way of doing science aimed predominantly at understanding of Nature and a way of doing science aimed at a control over Nature---an oscillation that, as we  have seen, has characterized science since antiquity---still occurs in contemporary physics.


\section{Between understanding and control:\\ How could science be different?}


In this essay we have addressed the question of how can science be different by looking at its history. We have proposed an analysis which identifies two souls of science: On the one hand, a side of science that aims at understanding Nature (more related to identifying metaphysical elements and their causal connections, and at putting forward theoretical hypotheses), and, on the other, the side that aims at controlling Nature (more related to empiricism, and technological applications). We have shown how science has progressed thanks to an alternation and an interplay between these two souls.

The distinction between explanatory science and understanding science here proposed resembles to some extent the structure of scientific development as  proposed by Kuhn  \cite{kuhn}. Namely, the alternation between “revolutionary science" (when new ideas thrive to explain observed anomalies) and “normal science" (characterized by puzzle solving in an established paradigm), which seem somehow related to the explanatory and the controlling natures of science, respectively. However, in our reconstruction, the proposed organisation of science is manifestly less linear. In Kuhn's view there is always a revolution after a period of normal science because that is the only mechanism for paradigm shift. In our view there is not necessarily a regular alternation of one period of understanding and one period of control over Nature. Simply, science has two complementary and yet not mutually exclusive souls: sometimes one prevails over the other, but they can coexist and perhaps they should coexist. 

Notice that we are not supporting the view that there should be a preference towards the exquisitely theoretical science, aimed at pure explanation without any applications. This is not only unrealistic and quite naive, but would also go against a large part of the motivations that animate human  search for better life standards. We live in a complex world and being able to control its phenomena in relation to important aspects of life   is essential to avoid to leave in fear. On the other hand, a science solely aimed at applications that leaves no room for bold theoretical conjectures, for speculation and in general for explanation would sooner or later reach its  limits and would take away from science the possibility of being one of the driving forces for new ideas (in the same way that philosophy, art, and literature  are). We stress again that this twofold character of science should not be regarded as a defect, a degeneration, a bias to be eliminated because it corrupts its genuine essence. On the contrary, this is an integral feature of the foundations of science. We have indeed seen several historical instances in which the two intrinsic natures of science had turned out to be complementary, also insofar as one may stimulate the growth of the other. 

Moreover, we should remark that the distinction between understanding and control of nature is hardly a deliberate choice of scientists. They usually try to replicate effects on demand when is possible, but it's often the case that scientists are forced to only deal with  events that cannot be locally reproduced (e.g., an event that would generate gravitational waves such as the explosion of a supernova). In such cases, one is  forced to stay at the explanatory level and  introduce theoretical and metaphysical terms that allow to tell a consistent story.
 
We hope to have brought some awareness concerning the fact that science is a cultural product to a large extend, and, as such, it replicates many aspect of the structure of society.
 This was shown through a quick historical overview (in fact, this essay covered an average of over 350 years per page), where many examples where presented to relate the development of a certain kind of science, with a focus on the predominance of one of its two intrinsic natures, in relation to the values, the ideas, the social demands of different epochs. Of course, we are exploring here the domain of counterfactual reasoning: The history of this world, and therefore of its science, is only one. But since it is possible to identify strong correlations between the  cultural contexts of different periods and the way science was perceived and produced, one can speculate that science could indeed have been different in different social context, and that  could be different in the future.
 
In this view, asking whether science could have been or could be different boils down to a large extent to asking: could have been the society of humans different? 
While according to our best current \textit{understanding} of Nature we cannot change the past, society could definitely be better in the future, which, if this essay contains anything sensible, would hopefully lead to a better structure of science too. 

How could science be better? By embracing its twofold, complementary nature. This requires a shift in the mindset, in the way we teach and communicate science, which, regrettably, far too often regards with indifference or even with reluctance philosophical thinking (and historical approaches) that could lead to novel understanding. This attitudes boils down to artificially extirpate half of the intrinsic nature of science. 
But awareness is always the first step towards evolution.

%


%



\begin{thebibliography}{0}%
\makeatletter
\providecommand \@ifxundefined [1]{%
 \@ifx{#1\undefined}
}%
\providecommand \@ifnum [1]{%
 \ifnum #1\expandafter \@firstoftwo
 \else \expandafter \@secondoftwo
 \fi
}%
\providecommand \@ifx [1]{%
 \ifx #1\expandafter \@firstoftwo
 \else \expandafter \@secondoftwo
 \fi
}%
\providecommand \natexlab [1]{#1}%
\providecommand \enquote  [1]{``#1''}%
\providecommand \bibnamefont  [1]{#1}%
\providecommand \bibfnamefont [1]{#1}%
\providecommand \citenamefont [1]{#1}%
\providecommand \href@noop [0]{\@secondoftwo}%
\providecommand \href [0]{\begingroup \@sanitize@url \@href}%
\providecommand \@href[1]{\@@startlink{#1}\@@href}%
\providecommand \@@href[1]{\endgroup#1\@@endlink}%
\providecommand \@sanitize@url [0]{\catcode `\\12\catcode `\$12\catcode
  `\&12\catcode `\#12\catcode `\^12\catcode `\_12\catcode `\%12\relax}%
\providecommand \@@startlink[1]{}%
\providecommand \@@endlink[0]{}%
\providecommand \url  [0]{\begingroup\@sanitize@url \@url }%
\providecommand \@url [1]{\endgroup\@href {#1}{\urlprefix }}%
\providecommand \urlprefix  [0]{URL }%
\providecommand \Eprint [0]{\href }%
\providecommand \doibase [0]{http://dx.doi.org/}%
\providecommand \selectlanguage [0]{\@gobble}%
\providecommand \bibinfo  [0]{\@secondoftwo}%
\providecommand \bibfield  [0]{\@secondoftwo}%
\providecommand \translation [1]{[#1]}%
\providecommand \BibitemOpen [0]{}%
\providecommand \bibitemStop [0]{}%
\providecommand \bibitemNoStop [0]{.\EOS\space}%
\providecommand \EOS [0]{\spacefactor3000\relax}%
\providecommand \BibitemShut  [1]{\csname bibitem#1\endcsname}%
\let\auto@bib@innerbib\@empty
\end{thebibliography}%


\begin{thebibliography}{200}


\bibitem{china} Needham J. \textit{et al.} (eds), 1954-ongoing. \textit{Science and civilization in China}, Cambridge University Press, Cambridge.


\bibitem{lindberg} Lindberg, D. C., 2010. \textit{The beginnings of Western science: The European scientific tradition in philosophical, religious, and institutional context, prehistory to AD 1450}. University of Chicago Press, Chicago.

\bibitem{curd} Curd, P. (ed.), 2011. \textit{A Presocratics Reader: Selected Fragments and Testimonia}. Hackett Publishing, Indianapolis/Cambridge.

\bibitem{rovelli} Rovelli, C., 2023. \textit{Anaximander: And the Birth of Science.} Riverhead Books, New York.


\bibitem{popper} Popper, K.R., 1958. Back to the Pre-Socratics: The presidential address. In \textit{Proceedings of the Aristotelian Society} (Vol. 59, pp. 1-24). Aristotelian Society, Wiley, New York.

\bibitem{russo} Russo, L., 2003. \textit{The forgotten revolution: How science was born in 300 BC and why it had to be reborn.} Springer Science and Business Media, Berlin.


\bibitem{rufini} Rufini, E., 1926. \textit{Il metodo di Archimede: E le origini del calcolo infinitesimale nell'antichità}, Nicola Zanichelli Editore, Bologna.


\bibitem{plutarch} Plutarch, \textit{The Life of Julius Caesar}. In \textit{The Parallel Lives}, Vol VII.


\bibitem{boyer} Boyer, C., 1968. \textit{A history of mathematics}. Wiley and Sons, New York.

\bibitem{hessen} Freudenthal, G. and McLaughlin, P. (eds.), 2009.\textit{ The social and economic roots of the scientific revolution: Texts by Boris Hessen and Henryk Grossmann} (Vol. 278). Springer Science and Business Media, Berlin.

\bibitem{fleck} Fleck, L., 2012. \textit{Genesis and development of a scientific fact.} University of Chicago Press, Chicago.

\bibitem{kuhn} Kuhn, T. S., 2012. \textit{The structure of scientific revolutions.} University of Chicago Press, Chicago.

\bibitem{merton} Merton, R. K., 1938. Science, technology and society in seventeenth century England. \textit{Osiris}, 4, pp.360-632.

\bibitem{merton2}  Merton, R. K., 1942.\textit{ The sociology of science: Theoretical and empirical investigations}, University of Chicago Press, Chicago.

\bibitem{sts} Felt, U., Fouché, R., Miller, C.A. and Smith-Doerr, L. (eds.), 2016. \textit{The handbook of science and technology studies.} MIT Press, Boston.


\bibitem{newton} Newton, I., Motte, A. and N. W. Chittenden (eds.), 1848, \textit{Newton's Principia. The mathematical principles of natural philosophy}.  D. Adee, New-York.




\bibitem{baracca} Baracca, A., 2021. \textit{Scientific Developments Connected with the Second Industrial Revolution:} A. Baracca, S. Ruffo, and A. Russo, \textit{Scienza e industria 1848–1915, 41 years later}.

\bibitem{maxwell} Maxwell, J. C., 1865. A dynamical theory of the electromagnetic field. \textit{Philosophical transactions of the Royal Society of London}, (155), pp.459-512.

\bibitem{lakatos} Lakatos, I., 1980. \textit{The Methodology of scientific research programmes.} Cambridge University Press, Cambridge.

\bibitem{darwin} Darwin, C., 1859. \textit{On the origin of species: By means of natural selection.} John Murray Press, London. 

\bibitem{letters} Darwin C. Letter to Asa Gray on Nov. 29th 1859 (79). In Darwin, F. (ed.), 1903. \textit{More letters of Charles Darwin.} John Murray Press, London. 





\bibitem{jammer} Jammer, M., 1966. \textit{The conceptual development of quantum mechanics.} McGraw-Hill, New York.

\bibitem{planck} Planck, M., 1949. \textit{Scientific Autobiography: And Other Papers}. Citadel Press, New York.



\bibitem{forman} Forman, P., 1971. Weimar culture, causality, and quantum theory, 1918-1927: Adaptation by German physicists and mathematicians to a hostile intellectual environment. \textit{Historical studies in the physical sciences}, 3, pp.1-115.





\bibitem{kaiser} Kaiser, D., 2011. \textit{How the hippies saved physics: Science, counterculture, and the quantum revival}. WW Norton and Company, New York.

\bibitem{bell} Bell, J. S., 1964. On the Einstein Podolsky Rosen paradox. \textit{Physics Physique Fizika}, 1(3), p.195.

\bibitem{freire} Freire Jr., O., 2014.\textit{ The quantum dissidents: Rebuilding the foundations of quantum mechanics (1950-1990)}. Springer Berlin, Heidelberg.

\bibitem{io} Del Santo, F., 2022. The foundations of quantum mechanics in post-war Italy’s cultural context, in Freire Jr., O. (ed.), \textit{The Oxford Handbook of the History of Quantum Interpretations}. Oxford University Press, Oxford.

\bibitem{ioemanu} Del Santo, F. and Schwarzhans, E., 2022. “Philosophysics” at the University of Vienna: The (pre-)history of foundations of quantum physics in the Viennese cultural context. \textit{Physics in Perspective,} 24(2-3), pp.125-153.



\bibitem{oxford} Freire Jr, O., Bacciagaluppi, G., Darrigol, O., Hartz, T., Joas, C., Kojevnikov, A. and Pessoa Jr, O. (eds.), 2022. \textit{The Oxford Handbook of the History of Quantum Interpretations.} Oxford University Press, Oxford.

\end{thebibliography}
\end{document}